\documentclass[aps,prl,superscriptaddress,showpacs]{revtex4-1}	
 \usepackage{amsmath}  
 \usepackage{makeidx}
 \usepackage{graphicx,color}
 \usepackage{amsfonts}
 \usepackage[ansinew]{inputenc}
 \usepackage[usenames,dvipsnames]{pstricks}
 \usepackage{subfigure}
 \usepackage{epsfig}
 \usepackage{pst-grad}
 \usepackage{pst-plot}
 \usepackage{mathrsfs}
 \def\vect#1{\mbox{\boldmath $#1$}}
	          \newcommand{\oo}{${{^{16}}{\rm O}}$}
              \newcommand{\nene}{${^{20}{\rm Ne}}$}
              \makeindex
\begin{document}
\title{Nonlocalized Clustering: A New Concept in Nuclear Cluster Structure Physics }
\author{Bo Zhou}
\email{zhoubo@rcnp.osaka-u.ac.jp.}
\email{bo.zhou@riken.jp.}
\affiliation{Department of Physics, Nanjing University, Nanjing 210093, China}
\affiliation{Research Center for Nuclear Physics (RCNP), Osaka University, Osaka 567-0047, Japan}
\affiliation{Nishina Center for Accelerator-Based Science, The Institute of Physical and Chemical Research (RIKEN), Wako 351-0198, Japan}
 \author{Y. Funaki}
 \email{funaki@riken.jp.}
\affiliation{Nishina Center for Accelerator-Based Science, The Institute of Physical and Chemical Research (RIKEN), Wako 351-0198, Japan}
 \author{H. Horiuchi}
 \affiliation {Research Center for Nuclear Physics (RCNP), Osaka University, Osaka  567-0047, Japan}
 \affiliation {International Institute for Advanced Studies, Kizugawa 619-0225,  Japan}	
\author{Zhongzhou Ren}
\email{zren@nju.edu.cn.}
\affiliation{Department of Physics, Nanjing University, Nanjing 210093, China}
\affiliation{Center of Theoretical Nuclear Physics, National Laboratory of Heavy-Ion Accelerator, Lanzhou 730000, China}  
\author{G.~R\"{o}pke}  
\affiliation {Institut f\"{u}r Physik, Universit\"{a}t Rostock, D-18051 Rostock, Germany}
\author{P. Schuck}
\affiliation{Institut de Physique Nucl\'{e}aire, Universit\'e Paris-Sud, IN2P3-CNRS, UMR 8608, F-91406, Orsay, France}
\affiliation{Laboratoire de Physique et Mod\'elisation des Milieux Condens\'es, CNRS-UMR 5493, F-38042 Grenoble Cedex 9, France}
 \author{A. Tohsaki}
 \affiliation{Research Center for Nuclear Physics (RCNP), Osaka University, Osaka 567-0047, Japan}
\author{Chang Xu}
\affiliation{Department of Physics, Nanjing University, Nanjing 210093, China}
\author{T. Yamada}
\affiliation{Laboratory of Physics, Kanto Gakuin University, Yokohama 236-8501, Japan}

\date{\today}

\begin{abstract}
We investigate the  $\alpha$+\oo\  cluster structure in the inversion-doublet band ($K^\pi=0_{1}^\pm$) states of \nene\  with an  angular-momentum-projected version of the Tohsaki-Horiuchi-Schuck-R\"{o}pke (THSR)  wave function, which was successful  ``in its original form" for the description of, e.g., the famous Hoyle state. In contrast with the traditional view on clusters as localized objects,  especially in inversion doublets, we find that these {\it single}  THSR wave functions, which are based on the concept of nonlocalized clustering, can well describe the $K^{\pi}=0_1^-$ band and the $K^{\pi}=0_1^+$ band. 
For instance, they have 99.98\% and 99.87\% squared overlaps for $1^-$ and $3^- $ states (99.29\%, 98.79\% and 97.75\% for  $0^+, 2^+$ and $4^+$ states), respectively, with the corresponding exact solution of the  $\alpha$+\oo\  resonating group method. These astounding results shed a completely new light on the physics of low energy nuclear cluster states in nuclei: The clusters are nonlocalized and move around in the whole nuclear volume, only avoiding  mutual overlap due to the Pauli blocking effect. 
\end{abstract}

\maketitle
The formation of clusters is a general problem in many-body physics that occurs in different systems such as 
ultracold gases in traps \cite{wenz} or the electron-hole-exciton system in excited semiconductors \cite{moskalenko} and other Coulomb systems, which also plays an important role in astrophysics \cite{rolfs}. In particular, it is one of the most important features in light nuclei.  Strongly correlated nucleons  can compose a cluster subunit, typically, the alpha 
cluster. Then the relative motion between clusters becomes a fundamental 
degree of freedom, as is the single-nucleon motion in the shell 
model  \cite{wildermuth}. This relative motion of clusters displays a complex 
feature due to  the character of the nuclear interaction and  the action of 
the Pauli principle, i.e., antisymmetrization. References. \cite{martinfreer2007,
vonoertzen2006, horiuchi_recent_2012, yamada_lnp} contain recent reviews 
on nuclear cluster physics.

The concept of localized clustering is a traditional understanding for the 
cluster structure in nuclei, which has a long history.  In 1937,  Wefelmeier 
\cite{wefelmeier_geometrisches_1937} proposed a cluster model  in which 
the  $\alpha $ nuclei could be considered as a collection of structureless rigid alpha 
particles undergoing localized motion.  Subsequently, Brink developed this 
idea and proposed  a microscopic cluster wave function \cite{brinkcluster} 
for  the cluster structure in nuclei.  The  geometrical  cluster structure  
was obtained by the energy variational method  without any prior assumptions 
\cite{brinkcluster,fujiwara_chapter_1980,rae_geometry_1994}.  However, only 
the superposition of Brink wave functions, that is the generator coordinate method (GCM) Brink wave function, is equivalent to the  resonating group method (RGM)  wave function which is
obtained by fully solving the intercluster motion of the cluster system rather than a single Brink wave function.  For details,  see Refs. \cite{saito, hh62}. This indicates 
already that a single Brink wave function which is characterized by the localized 
cluster correlation cannot describe the cluster structure sufficiently well.

Recently, the proposed Tohsaki-Horiuchi-Schuck-R\"{o}pke (THSR) wave function  \cite{Tohsaki2001-Alpha, 
PhysRevC.67.051306}, based on the concept of nonlocalized clustering, 
has brought a new  perspective to the Hoyle state. The Hoyle state is now 
considered to be an alpha condensate state \cite{freer2012,Chernykh2007, Kirsebom2012,Funaki2009-Concepts}, in which the three alpha clusters occupy the same $(0S)$ orbit and make a nonlocalized motion. 
On the other hand, also the 3 $\alpha$ RGM/GCM wave function of the 
ground state of ${}^{12}$C was found to have about 93\%  squared overlap with 
a {\it single} THSR wave function \cite{PhysRevC.67.051306}. Therefore, there arises 
the following important and fundamental question: Is this nonlocalized clustering the essential 
feature of  general cluster states in nuclei? Recently, we demonstrated by using a 
deformed THSR type of wave function~\cite{PTP.108.297} that the concept of nonlocalized 
clustering can be extended to the more compact cluster structures 
of the positive parity ground-state band in \nene~\cite{bo.new.2012}. 

On the other hand, in \nene\ there exists a negative-parity rotational
band with $K^\pi=0_1^-$ whose band head is the $1^- $ state at $E_x$=5.79 MeV \cite{tilley}.
This band has been interpreted as forming an inversion doublet rotational
band of $\alpha$+\oo\ clusters together with the ground-state band \cite{Horiuchi277}.
The existence of the inversion doublet bands has been regarded as a
clear  manifestation of the existence of the parity-violating intrinsic state with the $\alpha$+\oo\ structure which is nothing but the intrinsic state due to the $\alpha$+\oo\  localized clustering.  Thus the existence of the inversion doublet bands has been an important basis of the localized clustering picture which has been the longstanding concept of nuclear cluster physics.
In the present Letter, we, therefore, try to obtain a conclusive answer to the question whether  cluster motion in nuclei is localized or nonlocalized in studying the inversion doublet bands in \nene, especially the $K^\pi=0_1^-$ band.  Our study will definitely lead us to the conclusion that the latter case is realized in nuclear physics, thus stirring up the so far prevailing opinion that nuclear cluster phenomena are based on a rigid-body concept.
 
For our purpose, we first propose a hybrid wave function which includes the 
Brink wave function and the THSR wave function  as special cases. It is given by
\begin{equation*}
\Phi_{\rm cluster}(\vect{\beta}_i, \vect{S}_i) =\int d^3R_1\ldots d^3R_n   \times
\end{equation*}
\begin{equation}
\label{THSR1}
 \exp\{-\sum_{i=1}^{n}
(\frac{R_{ix}^2}{\beta_{ix}^2} +\frac{R_{iy}^2}{\beta_{iy}^2} +\frac{R_{iz}^2}{\beta_{iz}^2} ) \}                                 \Psi_{\rm cluster} ^B(\vect{R}_1+\vect{S}_1,\ldots, \vect{R}_n+\vect{S}_n)
\end{equation}
\begin{equation}
\label{THSR2}
\propto {\cal A}[\prod^{n}_{i=1}\exp\{-A_{i} \sum_{k=x,y,z}{\frac{( \vect{X}_{i}-\vect{S}_{i})_k^2}{B_{ik}^2}} \}\phi(C_{i}) ],
\end{equation}
where 
\begin{equation}
B^2_{ik}=2 b^2+ A_i\beta_{ik}^2, (k=x,y,z).
\end{equation}
Here $\vect{X}_i$  and $A_i$ are the center-of-mass coordinate and the mass number, respectively.  $\Psi_{\rm cluster} ^B(\vect{R}_1,\ldots,\vect{R}_n)$ is the Brink  wave function \cite{brinkcluster}.  $\phi(C_{i})$ is the internal wave function of the cluster $C_{i}$, which is usually described by the harmonic oscillator shell model wave function;  $b$  is the corresponding oscillator parameter.   

Compared with the original  THSR wave function, we introduce here another generator coordinate $\vect{S}_i$ in this new function Eq. (\ref{THSR1}). In this simple way, we find that this hybrid wave function Eq. (\ref{THSR2}) combines the important traits of the Brink model and the THSR model. When $\vect{S}_i$=0, Eq. (\ref{THSR2}) corresponds to the  THSR wave function and $\vect{\beta}_i$ [$\vect{\beta}_i  \equiv  (\beta_{ix}, \beta_{iy}, \beta_{iz})$] becomes the size parameter of the nucleus.  When $\vect{\beta}_i= 0$, this equation is nothing more but the original Brink wave function and  $\vect{S}_i$ represents the intercluster distance in nuclei.

Now, based on the above Eq. (\ref{THSR1}), the following cluster wave function of \nene\ can be obtained after some simplifications:
\begin{eqnarray}
&&\hspace{-1cm}\Psi_{\text{Ne}}(\vect{\beta},\vect{S}) \propto \exp(-\frac{10X_{G}^2}{b^2}) \nonumber \\
&&\times {\cal A}[\exp(-\sum_{k=x,y,z}\frac{8(\vect{r-S})_k^2}{5B_k^2}) \phi(\alpha)\phi({}^{16}\text{O})]. \label{ne2}
\end{eqnarray} 
Here $ \vect{\beta} \equiv (\beta_x, \beta_y, \beta_z)$, $B_k^2=b^2+2 \beta_k^2, (k=x,y,z)$,  $\vect{r}=\vect{X}_{1}-\vect{X}_{2}$, and $\vect{X}_G=(4\vect{X}_1+16\vect{X}_2)/20$. $\vect{X}_{1}$ and $\vect{X}_{2}$ represent the center-of-mass coordinates of the $\alpha$  and \oo\ clusters, respectively.  
All calculations here are performed with restriction to axially
symmetric deformation, that is, $\vect{S}\equiv (0,0,S_z)$.  Spin and parity eigenfunctions are obtained by the angular-momentum-projection technique; see below 
and Ref. \cite{bo.new.2012}.

\begin{figure}[!h]
\centering
\includegraphics[scale=0.37]{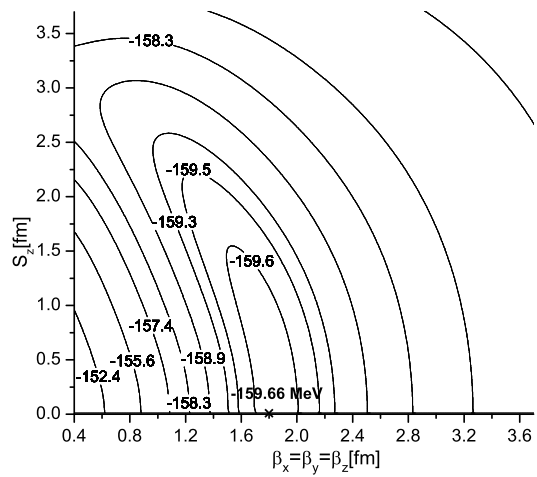}
\caption{\label{0+} Contour map of the energy surface of the  $J^\pi=0^+ $ state in the two-parameter space, $S_z$ and $\beta_{x}=\beta_{y}=\beta_{z}$.}
\end{figure}

\begin{figure}[!h]
\centering
\includegraphics[scale=0.37]{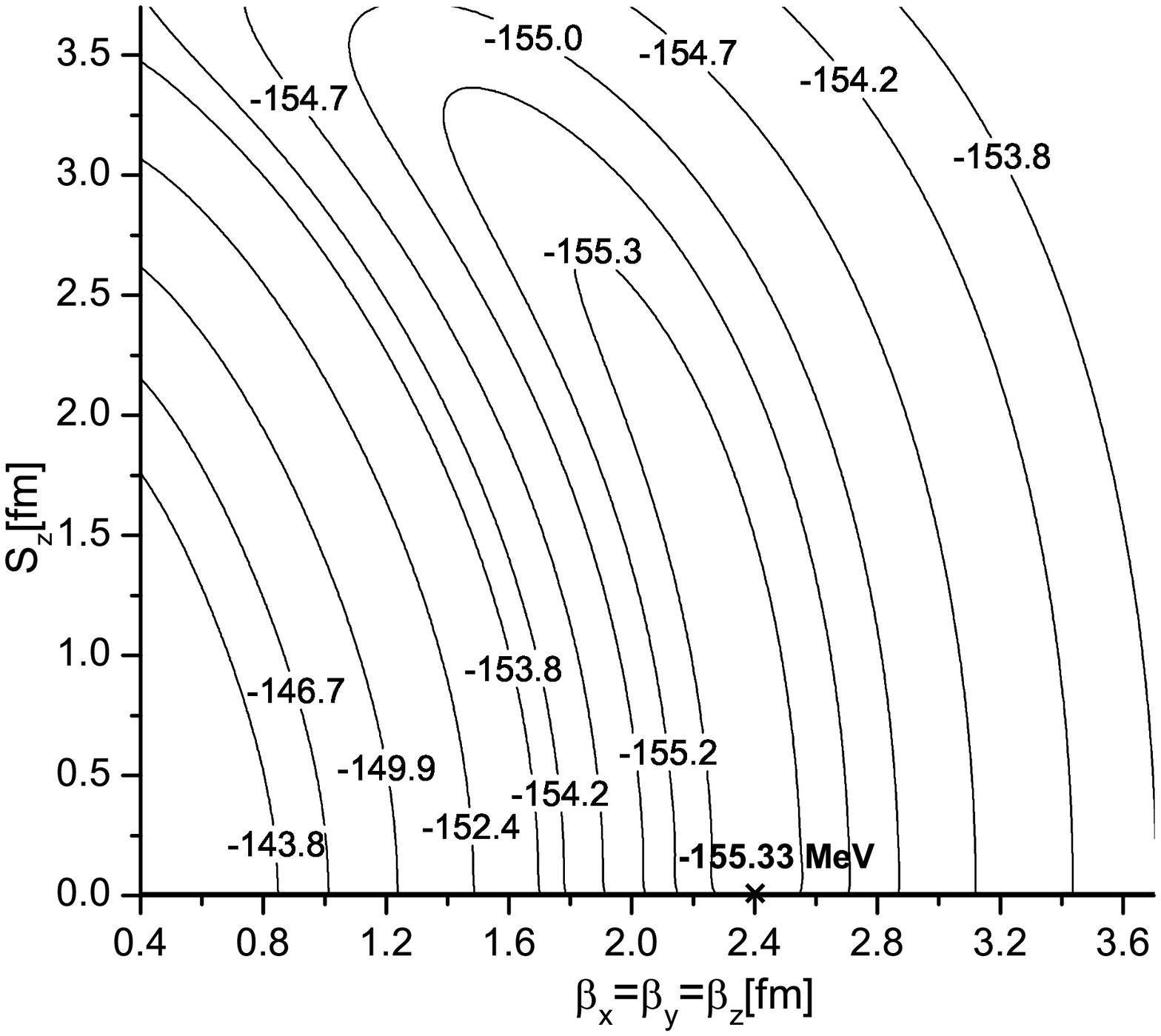}
\caption{\label{1-sz-bxyz} Contour map of the energy surface of the  $J^\pi=1^- $ state in the two-parameter space, $S_z$ and $\beta_{x}=\beta_{y}=\beta_{z}$.}
\end{figure}

\begin{figure}[!h]
\centering
\includegraphics[scale=0.37]{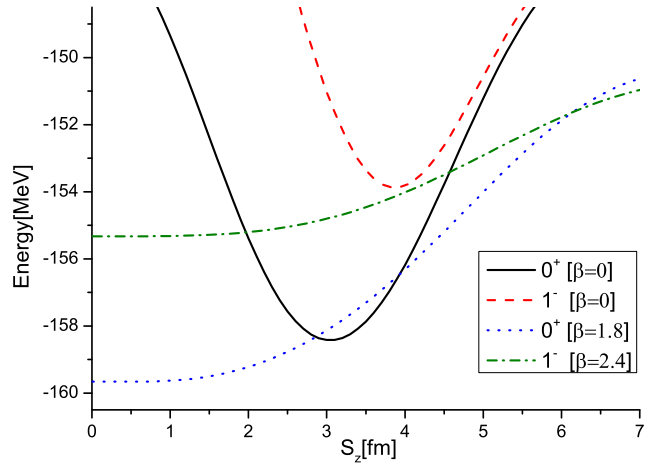}
\caption{ (Color online). Energy curves of $J^\pi=0^+$, $1^-$ states with different widths of Gaussian relative wave functions in the hybrid model.}
\label{nonlocal}  
\end{figure}

\begin{figure}[!h]
\centering
\includegraphics[scale=0.70]{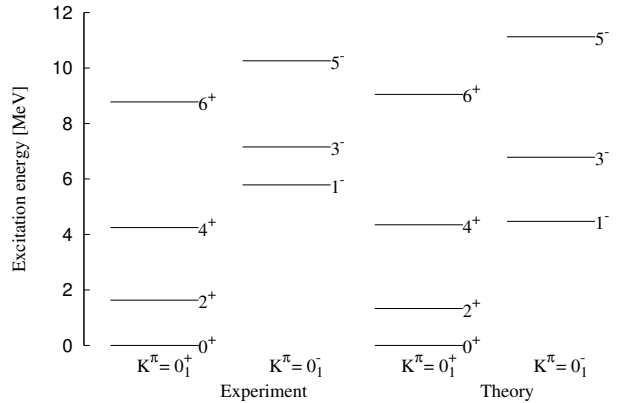}
\caption{\label{level} The energy levels of the inversion doublet bands in \nene\ reproduced by the hybrid wave function compared with the experimental levels.}
\end{figure}

\begin{table*}[htbp]
\centering
\caption{\label{table0} $E_{\text{Min}}^{\text{Brink}}(R)$  are the minimum energies at the corresponding $\alpha$-\oo\ distance $R$ in the Brink cluster model. $E_{\text{Min}}^{\text{Hyb}}(\beta_x, \beta_z)$ are the minimum energies at the corresponding values of $\beta_x=\beta_y$ and $\beta_z$ in the hybrid model. $E_{\text{GCM}}^{\text{Hyb}}$  are the GCM energies obtained by the hybrid model. We also show the squared overlaps between  our single normalized hybrid wave functions $ \hat{\Phi}_{\text{Min}}^{\text{Hyb}}$ corresponding to the minimum energies and the normalized  GCM Brink wave functions. The squared overlaps between  $ \hat{\Phi}_{\text{Min}}^{\text{Hyb}}$   and the single normalized Brink wave functions corresponding to their minimum energies are also listed. For the resonant state  $J^\pi=5^- $, we do not list the related GCM results. Units of energies are MeV.}
\begin{tabular}{ c c c c c c c c}
\hline
\hline
State  &$E_{\text{Min}}^{\text{Brink}}(R)$  & $E_{\text{Min}}^{\text{Hyb}}(\beta_x, \beta_z)$ & $E_{\text{GCM}}^{\text{Hyb}}(\text{Excited})$ & Experiment &   $|\langle\hat{\Phi}_{\text{Min}}^{\text{Hyb}}|\hat{\Phi}_{\text{Min}}^{\text{Brink}} \rangle|^2$ & $|\langle\hat{\Phi}_{\text{Min}}^{\text{Hyb}}|\hat{\Phi}_{\text{GCM}}^{\text{Brink}} \rangle|^2$  \\ \hline
$1^-$&$-153.87(3.9)$&$-155.38(3.7, 1.4)$&$-155.38(4.67)$&$-154.85(5.79)$&$0.9048$&$0.9998$ \\
$3^-$&$-151.40(3.8)$&$-153.07(3.7, 0.0)$&$-153.08(6.99)$&$-153.49(7.16)$&$0.8863$&$0.9987$ \\
$5^-$&$-146.81(3.6)$&$-148.72(3.3, 0.0)$& ----- &$-150.38(10.26)$& -----  & ----- \\
 \hline
 \hline
 \end{tabular}
\end{table*}

We now will calculate the energy as a function of the two parameters $\vect{\beta}$ 
and $S_z$. In this work,  the oscillator parameter $b$ is fixed at 1.46 fm, which is obtained by the variational calculation for the ground state of \nene. The Volkov no.1 force \cite{volkov1965equilibrium} is adopted as the nuclear interaction.
Figure \ref{0+}  shows the contour map of the energy surface of the  $J^\pi=0^+ $ state in the two-parameter space, $S_z$ and $\beta_{x}=\beta_{y}=\beta_{z}$. The minimum energy $-159.66$ MeV  appears at $S_z=0$ and $\beta_x=\beta_y=\beta_z=1.8$ fm.  For the other positive-parity states of the ground-state band in \nene, the minimum-energy points also appear at $S_z=0$.   Thus the hybrid wave functions are equal to the THSR wave functions in the description of these positive-parity states.

Figure \ref{1-sz-bxyz} shows the contour map of the energy surface of the  $J^\pi=1^- $ state in the two-parameter space, $S_z$ and $\beta_{x}=\beta_{y}=\beta_{z}$. We can see that the minimum energy  $-155.33$ MeV appears at $\beta_x=\beta_y=\beta_z=2.4$ fm and $S_z=0$. For the $J^\pi=3^-,5^-$ states, the minimum-energy points also appear at $S_z=0$ with different  $\vect{\beta}$ values. As we mentioned, the hybrid wave function 
without angular-momentum projection is equal to the THSR wave function at the limit of $S_z=0$, which is of positive parity. However, the hybrid wave function with odd angular momentum maintains its negative parity even in the limit of $S_z=0$, if it is normalized. This point can be explained as follows 
: The angular-momentum projection of $\Psi_{\text{Ne}}(\vect{\beta},\vect{S}) / \exp(-10X_G^2/b^2)$ gives us 
\begin{eqnarray}
&&\hspace{-1cm} {\cal A}\Big[j_L(2i \gamma S_z r)Y_{LM}(\widehat{r})e^{-\gamma r^2}\phi(\alpha)\phi(^{16}\text{O})\Big] \nonumber \\
&& \propto S_z^L \Phi_{LM}^0 + {\cal O}(S_z^{L+2}),\  \gamma=\frac{8}{5B^2}, \nonumber \\
&& \hspace{-1cm} \Phi_{LM}^0 = {\cal A}\Big[ r^L e^{-\gamma r^2} Y_{LM}(\widehat{r})\phi(\alpha) \phi(^{16}\text{O})\Big].
\label{angpro}
\end{eqnarray}
In Eq. (\ref{angpro}), the wave function is proportional to the parameter $S^L_z$. After normalization, this parameter drops out and the limit $S_z=0$ can be taken analytically. We, thus, obtain via the detour over the hybrid ansatz  $\Phi_{LM}^0$ as the THSR wave function having definite spin $L$ and parity independent of the localisation parameter $S_z$.

Of course, in practical calculations it is much easier for the 
reason given above to take for the parameter $S_z$   a very small value 
close to zero  within the limits of  numerical accuracy. We, therefore, 
see that the introduction of the parameter $S_z$ is just a very convenient 
way to introduce an angular-momentum-projected form of the THSR wave function of even {\it and} odd parity.  

On the other hand, this parameter now allows us to discuss more deeply the 
important issue of localization or nonlocalization of the clusters.
The competition between the two parameters $\vect{\beta}$ and $S_z$ leads to $S_z=0$. 
This indicates that in the typical case of \nene\ the clustering is nonlocalized.
Figure \ref{nonlocal} shows the energy curves of the $J^\pi=0^+$, $1^-$ states with different values for the widths of the Gaussian 
wave functions in the hybrid model. If $\vect{\beta}$ is fixed at 0, then the hybrid wave function becomes the  Brink wave function. In this case, $S_z$ is the intercluster distance parameter. For the ground state of \nene, the minimum energy appears at $S_z=3.0$ fm. For the $J^\pi=1^-$ state,  the optimum position appears at $S_z=3.9$ fm.  The nonzero values of $S_z$ seem to indicate that  the $\alpha$+\oo\ structure of \nene\  favors localized clustering. This is just the traditional concept.  However, now we think this argument is misleading.  The nonzero minimum point $S_z$  simply occurs because the 
width of the Gaussian wave function of the relative motion in the Brink model 
is fixed to a narrow wave packet. If we take for $ \beta_x=\beta_y=\beta_z $=$1.8$  and $2.4$ fm for $J^\pi=0^+$, $1^-$, respectively, according to the minimum positions in Fig. \ref{0+}  and \ref{1-sz-bxyz}, namely, if we use a broad enough width of  Gaussian wave packet for describing the relative motion, then, we find that the minimum points appear at $S_z=0$  in Fig. \ref{nonlocal}. 
This indicates that the separation distance parameter $S_z$ does not play any physical role in describing the $\alpha$+\oo\ cluster structure, even for the negative-parity states, which have been considered to be the typical example of localized clustering.  Instead of that, the new parametrization by $\vect{\beta}$, which characterizes nonlocalized clustering, is most appropriate for describing the cluster structure in \nene. 

In Ref. \cite{bo.new.2012}, we have demonstrated for the ground-state band in \nene\ that the THSR wave functions at the minimum-energy points are almost $100$\% equivalent to the superposed Brink wave functions obtained by GCM calculations.  Their squared overlaps are 99.29\%, 98.79\%, and 97.75\%  for the $0^+$, $2^+$,  and $4^+$ states, respectively.  The result of the calculation in this Letter is that also for the $K^\pi=0_{1}^-$ band states the hybrid wave functions at the minimum-energy points are almost 100\% equivalent to the superposed Brink wave functions obtained by GCM calculations. We give in Table \ref{table0} the squared overlap values between them, which are $99.98$\% and $99.87$\% for $1^-$ and $3^-$ states, respectively. These results mean that a {\it single} THSR wave function is equivalent to the exact RGM solution of the $\alpha$+\oo\ two-body problem. This necessarily leads us to the new concept that $\alpha$ and \oo\ clusters move in a nonlocalized way, rather than the longstanding concept of localized, rigid-body-like clustering with a certain separation distance between the clusters.  This important conclusion was not possible with our work in Ref. \cite{bo.new.2012} when only the positive-parity states were considered.

Figure \ref{level} shows the energy levels of  the $K^\pi=0_{1}^-$ band reproduced by the hybrid wave function, together with the ground-state band  $K^\pi=0_{1}^+$ in \nene. Since it is well known that the calculation with GCM Brink wave function reproduced the two bands with $K^\pi=0_{1}^\pm$, it is then natural that we 
have good reproduction of experiments also with our {\it single} angular-momentum-projected THSR wave functions. It should be noted that the results are obtained without any adjustable parameter. So, the good 
agreement with the experimental values means that this THSR-type wave 
function grasps very well the character of the relative motion of the system.

But how shall we understand the rotational band based on the concept of nonlocalized clustering?
In the Brink model, because the $\alpha$-\oo\ Brink wave 
functions are deformed and parity-violating wave functions, it is easy to 
understand the reason why they can describe the inversion doublet rotational 
bands with $K^\pi=0_{1}^\pm$. On the other hand, in the case of the THSR wave 
function with $S_z=0$, i.e. with good symmetry, that is a wave function in 
the laboratory frame, the $\alpha$ and \oo\ clusters undergo a nonlocalized 
relative motion in the whole nuclear volume besides the volume of mutual overlap where their probability of presence 
is strongly reduced. Hence, it may
seem that the angular-momentum-projected wave function, i.e., with $S_z=0$, 
are not directly related with the $K^\pi=0_{1}^\pm$ inversion doublet bands. 
In fact, the strongly anisotropic $\beta_x=\beta_y \ne \beta_z$ values shown in Table~\ref{table0} indicate an oblate shape. On the other hand, in Ref.~\cite{bo.new.2012}, we showed that for the ground-state band, the $\beta_x=\beta_y \ne \beta_z$ values giving the minimum energies indicate a prolately deformed shape. 
However, for example, the oblately deformed hybrid THSR wave function with $\beta_x=\beta_y=3.7$ fm, $\beta_z=1.4$ fm, and $S_z = 0$ fm giving the minimum energy for the $1^-$ state has $99.98$\% squared overlap with the $1^-$ wave function projected from the prolately deformed THSR wave function with $\beta_x=\beta_y=0.1$ fm, $\beta_z=3.2$ fm, and $S_z = 0$ fm. This means that it is then possible to consider the prolately deformed THSR wave function as the intrinsic wave function, which can generate the $K^\pi=0_1^\pm$ inversion doublet bands.
A more detailed description how, in spite of the delocalized motion inherent to the THSR wave function, we have succeeded to obtain the inversion doublet bands ($K^{\pi}=0_1^+$ and $K^{\pi}=0_1^-$ bands with a small energy gap between the two bands) which are generally understood as a clear signature of an $\alpha$+\oo\ cluster structure in \nene\ is the subject of a forthcoming paper.  In short, the reason which will be given in that paper is that the intercluster Pauli repulsion (remember that the THSR wave function is fully antisymmetric) hinders  the clusters from coming too close to one another in their otherwise delocalized motion. In the special case of only two clusters which is the case here, this creates, nonetheless, an $\alpha$+\oo\ structure which, however, should not be mixed up with a rigid-body, dumbbell-like shape. 

In summary,  we have succeeded in obtaining a conclusive answer to the
question whether cluster motion in nuclei is localized or nonlocalized by
studying the inversion doublet bands in \nene.  Since the inversion doublet
bands in \nene\ have, in the past, been understood as a clear signature of localized cluster
structure of $\alpha$ + \oo,  it is necessary that the inversion doublet bands
are also well described by the wave functions of nonlocalized clustering, 
in order for the picture of the nonlocalized clustering to be adopted.  
We first showed by introducing a hybrid Brink-THSR wave function that
the energy minimum is obtained by a pure THSR wave function with $S_z$ = 0,
which shows that the energy minimum point obtained by a pure Brink wave
function with nonzero  $S_z$ does not support localized clustering. 
We further found the highly surprising fact that those single THSR
wave functions at the energy minimum points are nearly 100\% equivalent to
the exact RGM solution of the $\alpha$ + \oo\ system which is equivalent to the
superposed Brink wave functions obtained by the GCM.  These results mean
that the concept of nonlocalized clustering proposed by a THSR-type wave function
is essential in correctly understanding the cluster structures in nuclei, which is much
different from the localized clustering picture typically described by Brink
wave functions superposed only around the energy minimum point of the Brink
energy curve. These astonishing features revealed by the THSR-type wave functions
force us to adopt a completely new understanding of nuclear cluster physics in
the way outlined above.  

This work is supported by the National Natural Science Foundation of China (Grants No. 11035001, No. 10975072, and No. 11175085), by the 973 Program of China (2013CB834400), and by the Project Funded by the Priority  Academic Program Development of Jiangsu Higher Education Institutions (PAPD).

\end{document}